\documentclass[aps, preprint]{revtex4-2}
\usepackage{graphicx}
\usepackage{amsfonts}
\usepackage{amssymb}
\usepackage{amsbsy}
\usepackage{amsmath}
\usepackage{mathrsfs}
\usepackage{latexsym}
\usepackage{bm}
\usepackage{color}
\usepackage{xcolor}
\usepackage{wasysym}
\usepackage{hyperref}
\usepackage{physics}
\usepackage{cancel}
\usepackage{float}
\usepackage{bm}

\begin{document}
\sloppy

\title{Torsional four-fermion interaction and the Raychaudhuri equation}

\author{Shibendu Gupta Choudhury}
\email{shibendu17@bose.res.in}
\author{Sagar Kumar Maity}
\email{sagar.physics1729@bose.res.in}
\author{Amitabha Lahiri}
\email{amitabha@bose.res.in}

\affiliation{S. N. Bose National Centre for Basic Sciences,\\
JD Block, Sector 3, Salt Lake, WB 700106, INDIA.}


\date{\today}

\begin{abstract}

Intrinsic spin of fermions can generate torsion in spacetime. This torsion is a non-propagating field that can be integrated out, leaving an effective non-universal four-fermion interaction. This geometrical interaction affects fermions inside a matter distribution and can be expected to become stronger as the density grows. In this work, we investigate the role of this interaction in a gravitationally collapsing fermionic distribution. Our specific aim is to explore if this interaction can provide a repulsive contribution and prevent the final singularity formation. We consider a collapsing distribution of massive fermions, ignoring other interactions. Using simplified yet reasonable assumptions, we establish that a repulsive contribution can arise depending on how torsion couples with different chiralities. Also, the interaction starts to dominate as the collapse proceeds, accelerating or decelerating the collapse depending on the relative signs of the geometrical interaction between different species of fermions.

\end{abstract}

\maketitle
\newpage

\section{Introduction}
The final stage of the gravitational collapse of a compact object due to its own gravity is a major concern in the fields of relativistic astrophysics and gravitational physics.
It is generally believed that when gravity overpowers the pressure inside the compact object, the collapsing object undergoes a continuous contraction, leading to ever-increasing densities and curvatures. Eventually, this collapse process culminates in the formation of a curvature singularity in the vicinity of which densities, spacetime curvatures, and all other physical quantities become infinite, causing the framework of the underlying gravity theory to break down. For a comprehensive discussion in this regard we refer to~\cite{Joshi:2000fk, Joshi:2013coa, Joshi:2008zz, Joshi:2012mk, Malafarina2016} and references therein. 
In the context of General Relativity (GR), the Hawking and Penrose singularity theorems~\cite{Penrose:1964wq, Hawking:1970zqf, hawking_ellis_1973} predict that under physically reasonable conditions spacetime singularities are inevitable for purely gravitational systems (for a recent counterargument, see~\cite{Kerr:2023rpn}). A complete physical description of the spacetime structure thus demands resolution of these singularities. A widely accepted possibility is that the behaviour of such extreme regions may not be governed by the classical GR theory itself, and a quantum theory of gravity would be the most likely description. Another possibility is to consider other realistic gravitational theories whose geometrical attributes (not present in GR) may affect the final asymptotic stages of the collapse.

As the collapse proceeds, a massive star may go through stages in which it can be thought of as a fermionic system (e.g., white dwarf and neutron stars). When fermions couple with the spacetime structure, they induce torsion through their intrinsic spin. The Einstein-Cartan-Sciama-Kibble (ECSK) theory~\cite{Cartan:1923zea, Cartan:1924yea, Kibble:1961ba, Sciama:1964wt, Sciama:1964wt, Hehl:1976kj, Hehl:1974cn, Hammond:2002rm, Hehl:2007bn, Poplawski:2009fb} of spacetime is motivated by the desire to provide a simple description of 
the dynamics of fermions in curved spacetime. The most convenient way to do this is to write GR in a first-order formalism using tetrad fields or vierbeins and a spin connection as independent field variables. The field equations  of ECSK theory relate certain combinations of the curvature and torsion tensors to the energy-momentum and spin density tensors, respectively. Thus in the framework of ECSK theory, both mass and spin, which are intrinsic and fundamental properties of matter fields, affect the spacetime structure.

Since the advent of ECSK theory, various investigations have been carried out in the cosmological as well as astrophysical context addressing the  
resolution of the problem of spacetime singularities~\cite{TRAUTMAN1973, Arkuszewski:1975fz, Kerlick:1975tr, Stewart:1973ux, Hehl:1976kj}. The concept of a spin fluid, namely, Weyssenhoff fluid~\cite{Weyssenhoff:1947iua} is used in~\cite{Arkuszewski:1975fz} to show the avoidance of the final singularity in the 
collapse of such a spherical fluid distribution. 
Cosmological models in ECSK theory have been investigated to show that spacetime torsion may provide a framework by virtue of which the initial singularity of the Universe is replaced by a nonsingular bounce~\cite{Gasperini:1998eb, Dolan:2009ni, Magueijo:2012ug, Bambi:2014uua, Ziaie:2013pma, Huang:2015zma}. Work along this line has been also extended to study the effects of spin in the early universe~\cite{KOPCZYNSKI1972219, Kuchowicz1978, PhysRevLett.56.2873}.

When the torsion arises solely from the intrinsic spins of the fermions, the former can be characterized completely in terms of the fermionic fields. The fermions in turn experience an effective four-fermion interaction which affects the propagation of fermions through matter~\cite{Hehl:1976kj, 10.1063/1.1665738, 10.1063/1.1703678, Gursey1957, Gasperini:2017ggf, Chakrabarty:2019cau}.
In such a scenario, we can describe the dynamics of fermions using a modified energy-momentum tensor within the purview of GR, as 
we will discuss in detail in Sec.~\ref{sec2}. The geometrical four-fermion interaction can be significant when dense fermionic matter 
is present and thus it can have intriguing implications in the collapse of dense
configurations and also in the early Universe. It has been { argued} in~\cite{Kerlick:1975tr}  that this effective four-fermion interaction, albeit with only one species of fermions, could actually worsen the problem of singularity for the case of a Bianchi type I spacetime. On the other hand, {works in ~\cite{Dolan:2009ni, Bambi:2014uua, Alexander:2008vt} suggest}  that this interaction can mitigate and in some cases can eliminate the Big Bang singularity in the context of the Friedmann-Lemaître-Robertson-Walker (FLRW) cosmology.

In this work, we aim to examine how this effective interaction affects a gravitational collapse of a fermionic distribution. We will assume that the geometrical interaction couples right- and left-handed fermions with different strengths. For the complete picture of the corresponding collapse scenario, we need to solve the field equations and the nonlinear Dirac equation (Eqs.~\eqref{femain} and \eqref{ndceq} respectively) in the dynamical situations. This is actually a formidable task for a general system. Therefore, we need to consider simplified scenarios while keeping the essential physics. We shall consider a hot, large and dilute distribution for our study.
In order to carry out our investigation, we will adopt the covariant approach of using the Raychaudhuri equation~\cite{Raychaudhuri:1953yv, Ehlers1993, Kar2007, Ellis2007}.
In this approach, the worldlines of matter particles within
the collapsing distribution are considered  as the congruences under study. The occurrence (or avoidance) of a singularity is then characterized using the focusing (or defocusing) of the congruence. Germani and Tsagas~\cite{Germani:2005ar} introduced the concept of utilizing the Raychaudhuri equation to study a magnetized Tolman–Bondi collapse. Subsequently, this method has been employed in several other gravitational collapse models (see~\cite{Germani:2005ar, Tsagas:2006sh, Kouretsis:2010nu, Tsagas:2020lal, Choudhury:2019snb, Choudhury:2022uek} and references therein). Hensh and Liberati~\cite{Hensh:2021oyk} studied the Raychaudhuri equation and the Oppenheimer--Snyder type gravitational collapse model in the context of ECSK theory. They considered the presence of an independent background torsion field and a Weyssenhoff dust fluid for this purpose and showed that the formation of the singularity can be avoided. Our work has a different starting point. As already mentioned, we do not assume any background torsion -- the torsion here is induced by the dynamics of fermionic particles. Further, the torsion is non-dynamical and can be eliminated from the action. In addition, we will use the well-known Raychaudhuri equation in the Riemannian geometry and carry out the investigation under the  framework of GR, though with an energy-momentum tensor which includes the contribution of the effective geometrical interaction.

The paper is organized as follows. In Sec.~\ref{sec2}, we discuss how, for fermionic systems, starting from the ECSK formalism we can write an equivalent formalism where the underlying theory can be considered as GR with a modified energy-momentum tensor. Sec.~\ref{sec3} briefly discusses the Raychaudhuri equation and focusing of a congruence. The corresponding concepts for fermionic congruences is also discussed in this section. In Sec.~\ref{sec4}, the energy conditions are inferred using a few simplifying assumptions, and thus we comment on the fate of the focusing condition for the system under consideration. This determines if the effective four-fermion interaction helps or prevents focusing and hence the final singularity formation. 
Finally, we summarize our findings and address open questions in Sec.~\ref{sec5}. { We state the important results in the main text and put the details of the necessary calculations in the appendix.}


\section{Energy-Momentum tensor of the collapsing fermionic distribution}\label{sec2}
Fermions in curved spacetimes are most conveniently dealt with using the ECSK formalism. In this, the $\gamma$ matrices are taken to be on the tangent space of the manifold at each point, which is isomorphic to flat Minkowski space. We will indicate the tangent space objects by uppercase Latin indices and spacetime objects by lowercase Greek indices. The two kinds of objects are related by 
the mathematical construct of the vierbein or tetrads $e^\mu_I$, a set of four contravariant vector fields satisfying the orthonormality condition~\cite{Gasperini:2017ggf, Mielke:2017nwt, Chakrabarty:2018ybk}
\begin{equation}\label{metrics}
    g_{\mu\nu} e^\mu_{I} e^\nu_{J} = \eta_{IJ}\,, \qquad \eta_{IJ} e^I_\mu e^J_\nu = g_{\mu\nu}\,,
\end{equation}
where $g_{\mu\nu}$ is the spacetime metric and $\eta_{IJ}$ is the Minkowski metric $\eta_{IJ} = \text{diag}(-1, +1, +1, +1)$. 
Spacetime and Lorentz indices are raised and lowered with the spacetime and Minkowski metrics respectively. We have also defined the ``co-tetrad" 
fields $e_\mu^I$ as the inverse of the tetrad fields, 
\begin{equation}\label{tetrads}
e^\mu_I e_\nu^I = \delta^\mu_\nu\,, \qquad e^\mu_I e^J_\mu = \delta^J_I\,.
\end{equation}

In ECSK theory, the action for Dirac fields coupled to gravity is
%
%
\begin{equation} \label{action}
   S=\int|e| d^4 x \, \left[ \frac{1}{2}e^\mu_I e^\nu_J F_{\mu\nu}^{IJ}[e, A] + \frac{1}{2}\left(-\bar{\psi} \gamma^K e_K^\mu D_\mu \psi-\left(\bar{\psi} \gamma^K e_K^\mu  D_\mu \psi\right)^{\dagger}\right)- m \bar{\psi} \psi\right].
\end{equation}
Here $|e|$ is the determinant of the co-tetrad $e_\mu^I$ treated as a matrix $(|e| = \sqrt{|g|})$, $A_\mu^{IJ} = - A_\mu^{JI}$ are the components of the spin connection, while
$F = dA + A\wedge A $ is the curvature of the spin connection $A$. The condition that the connection is tetrad-compatible, $\nabla_\mu e_\lambda^I = 0\,,$ is sometimes referred to in the literature as the tetrad postulate. Using this, we can write the components of the affine connection as 
\begin{equation}
	\Gamma^\lambda_{\mu\nu} = e^\lambda_I \partial_\mu e^I_\nu + A^I_{\mu J} e^J_\nu e^\lambda_I\,.
\end{equation} 
Then $e^\mu_I e^\nu_J F_{\mu\nu}^{IJ}[e, A] = R(\Gamma)\,,$ the Ricci scalar corresponding to $\Gamma$.
The covariant derivative of a spinor $\psi$ is defined with the help of the spin connection as ${D}_\mu \psi = \partial_\mu \psi - \frac{i}{4} A_\mu^{IJ} \sigma_{IJ} \psi$, where $\sigma_{IJ}=\frac{i}{2}\left[\gamma_I,\gamma_J\right]_-\,.$
As we will see below, the above action carries torsion degrees of freedom which originate solely due to the presence of fermions.

Extremising the action~(\ref{action}) with respect to the spin connection $A_\mu^{IJ}$, we get,
\begin{equation}\label{tspcon}
     A_\mu^{I J}=\omega_\mu^{I J}\left[e\right]+\frac{i}{16} \bar{\psi}\left[\gamma_K, \sigma^{I J}\right]_+ \psi e_{\mu}^K,
\end{equation}
where $[a, b]_+ = ab +ba\,$ and $\omega_{\mu}^{IJ}$ is the torsion-free spin connection, which is related to the Christoffel symbols by 
\begin{equation}
 \omega_\mu^{IJ} =  \tilde{\Gamma}^\nu_{\sigma \mu}e_\nu^I e^{\sigma J} + e_\nu^I \partial_\mu e^{\nu J}.
 \label{torsion-free}
\end{equation}
We note that using Eq.~(\ref{metrics}) and the definition of Christoffel symbols, we can write $\omega_\mu^{IJ}$ fully in terms of the tetrads, 
co-tetrads, and their derivatives. The last term of Eq.~\eqref{tspcon} corresponds to the contorsion tensor. 
If we put back the solution Eq.~\eqref{tspcon} into the action~\eqref{action}, we can completely separate the torsion degree of freedom and write it as an effective interaction term, 
\begin{equation}\label{faction}
\begin{split}
    S[e, \omega, \phi, \psi]= \int|e| d^4 x \, \Bigg[\frac{1}{2}e^\mu_I e^\nu_J \tilde{F}_{\mu\nu}^{IJ}[e, \omega] + \frac{1}{2}\left(-\bar{\psi} \gamma^\mu  \Tilde{D}_\mu \psi-\left(\bar{\psi} \gamma^\mu  \Tilde{D}_\mu \psi\right)^{\dagger}\right)\qquad \\  - m \bar{\psi} \psi 
    - \frac{3}{16} (\Bar{\psi}\gamma_I\gamma_5\psi)(\Bar{\psi}\gamma^I\gamma_5\psi)\Bigg].
\end{split}
\end{equation}
Here the 
tilde signifies that we are working with the torsion-free spin connection and we have written $ \gamma^K e_K^\mu = \gamma^\mu\,.$

Now extremising the action with respect to the vierbein $e_\mu^I$ leads to the Einstein equations,
\begin{equation}\label{femain}
    \Tilde{R}_{\mu\nu} - \frac{1}{2} \Tilde{R} g_{\mu\nu} =  T_{\mu\nu},
\end{equation}
where the energy-momentum tensor $T_{\mu\nu}$ is obtained by extremising the matter part of the action (expression within the second integral in equation~\eqref{faction}) with respect to the tetrads,
\begin{equation}\label{eemt}
       T_{\mu\nu} = 
      {   \frac{1}{4} \sum_{f}^{}\Big[\Bar{\psi}_f \gamma_{(\mu} \Tilde{D}_{\nu)} \psi_f + \text{h.c.}\Big]+\frac{3}{16}g_{\mu\nu} (\Bar{\psi}\gamma_I\gamma_5\psi)(\Bar{\psi}\gamma^I\gamma_5\psi)\,.}
\end{equation}
On the other hand, extremising the action with respect to the field $\bar{\psi}$ yields the equation of motion of the spinor, a generalization of the Dirac equation to curved spacetime~\cite{10.1063/1.1665738, Hehl:1976kj},
\begin{equation}\label{nlde}
     \gamma^\mu \Tilde{D}_\mu \psi + m\psi = -  \frac{3}{8} \Bar{\psi} \gamma^I \gamma_5 \psi \gamma_I \gamma_5 \psi.
\end{equation}
Therefore, we can now consider the geometry to be Riemannian and work with simple General Relativity. The torsion due to the presence of fermions is now effectively contained in the matter sector in the guise of the interaction term in the energy-momentum tensor.

A different way of writing the action of Eq.~(\ref{action}) provides an way to include a more general interaction~\cite{Chakrabarty:2019cau}. If we start by separating out the torsion-free part of the spin connection as $A_\mu^{IJ} = \omega_\mu^{IJ} + \Lambda_\mu^{IJ} $\,, where $\omega_\mu^{IJ}$ is given by Eq.~(\ref{torsion-free}), we can treat $\Lambda_\mu^{IJ}$ as an independent field (this field is called contorsion in the literature). Then it is easy to see from varying the action that this $\Lambda$ is a non-dynamical field, with the solution
\begin{equation}
	\Lambda_\mu^{IJ}=\frac{i}{16} \bar{\psi}\left[\gamma_K, \sigma^{I J}\right]_+ \psi e_{\mu}^K\,.
	\label{contorsion.1}
\end{equation}
But we have the possibility of a more interesting interaction ---  the contorsion $\Lambda_\mu^{IJ}$ can couple with different species of fermions, as well as and  left- and right-handed fermions with different coupling strengths~\cite{Chakrabarty:2019cau}.
In that case, $\Lambda_\mu^{IJ}$ is again a non-dynamical field, so it can be replaced by its solution in the action, leading to a four-fermion interaction term
\begin{equation}
 -\frac{3}{16}\left[\sum_{f}^{}\left(-\lambda_{fL} \Bar{\psi}_{fL}\gamma_I\psi_{fL}+\lambda_{fR}\Bar{\psi}_{fR}\gamma_I\psi_{fR}\right)\right]^2\,,
 %
\label{four-fermion.1}
 \end{equation}
where the sum runs over all species of fermions. Here we have adopted the  notation $\left(P_I\right)^2\equiv P^I P_I$\,, where $P^I$ denotes any vector field. This interaction term can also be written in terms of vector and axial currents as
\begin{equation}
	\begin{split}
 {-\frac{1}{2}\left[\sum_f\left(\lambda_{fV} \Bar{\psi}_f\gamma_I\psi_f+\lambda_{fA} \Bar{\psi}_f\gamma_I\gamma_5\psi_f\right)\right]^2,}
\end{split}
\end{equation}
where we have written $\lambda_{V,A}=\sqrt{\frac{3}{4}}\left(\lambda_R\pm \lambda_L\right)$\,.
Consequently, the effective energy-momentum tensor is given by (see appendix~\ref{appb} for the calculation)
\begin{equation}\label{emt}
  T_{\mu\nu} =
 {   \frac{1}{4} \sum_{f}^{}\Big[\Bar{\psi}_f \gamma_{(\mu} \Tilde{D}_{\nu)} \psi_f + \text{h.c.}\Big] +\frac{1}{2}g_{\mu\nu}\left[\sum_f\left(\lambda_{fV} \Bar{\psi}_f\gamma_I\psi_f+\lambda_{fA} \Bar{\psi}_f\gamma_I\gamma_5\psi_f\right)\right]^2,}
\end{equation}
and the Dirac equation takes the form
\begin{equation}\label{ndceq}
      \gamma^\mu \Tilde{D}_\mu \psi_f + m\psi_f = - \left(\lambda_{fV}\gamma^I\psi_f+\lambda_{fA}\gamma^I\gamma_5\psi_f\right)\sum_{f'}^{}\left(\lambda_{f'V} \Bar{\psi}_{f'}\gamma_I\psi_{f'}+\lambda_{f'A}\Bar{\psi}_{f'}\gamma_I\gamma_5\psi_{f'}\right)\,.
\end{equation}
We will now explore the implications of this four-fermion interaction in the gravitational collapse of a fermionic distribution.  
As discussed earlier, we can work with a torsion-free connection but with an energy momentum tensor modified
by the four-fermion interaction term. What we are interested in is to see whether the interaction term will deter or assist 
gravitational collapse of a ferionic gas. We will use the approach based on the Raychaudhuri equation~\cite{Germani:2005ar, 
	Tsagas:2006sh, Kouretsis:2010nu, Tsagas:2020lal, Choudhury:2019snb, Choudhury:2022uek}, which we now discuss briefly.


\section{Raychaudhuri equation and Focusing}\label{sec3}
For a timelike geodesic congruence having velocity vector $u^\mu$, with $u_\mu u^\mu = -1$, the Raychaudhuri equation is given by
\begin{equation}\label{rceq}
  u^\mu \tilde{\nabla}_\mu \tilde{\theta}= \frac{ \mathrm{d}{\Tilde{\theta}}}{{\mathrm{d}\tau} }= - \frac{1}{3} \Tilde{\theta}^2 - \tilde{\sigma}_{\mu\nu}\tilde{\sigma}^{\mu\nu}
 +\tilde{\omega}_{\mu\nu}\tilde{\omega}^{\mu\nu} - \Tilde{{R}}_{\mu\nu} u^{\mu} u^{\nu}\,,
\end{equation}
where $\tilde{\theta}=\tilde{\nabla}_\mu u^\mu$ is the expansion scalar, $\tau$ is proper time, 
$\tilde{\sigma}_{\mu\nu}=\tilde{\nabla}_{(\nu}u_{\mu)}-\frac{1}{3}h_{\mu\nu}\tilde{\theta}$
is the shear tensor with $h_{\mu\nu} = g_{\mu\nu} + u_{\mu}u_{\nu}$ being the induced spatial metric, $\tilde{\omega}_{\mu\nu}=\tilde{\nabla}_{[\nu}u_{\mu]}$
is the rotation tensor and $\tilde{{R}}_{\mu\nu}$ is the Ricci tensor. The Raychaudhuri equation is a geometrical identity in Riemannian geometry and it does not need any gravitational theory to begin with. {As before, we denote with a tilde quantities calculated with a torsion-free connection.}

Usually it is the evolution equation for the expansion scalar, namely Eq.~\eqref{rceq}, which is primarily referred to as the Raychaudhuri equation and is used to study different phenomena, e.g. focusing of congruences, gravitational collapse, cosmological evolution, etc. However, there are two other equations which dictate the evolution of shear and rotation. The  evolution equation for shear is given by~\cite{Poisson:2009pwt, Wald:1984rg}
\begin{equation}\label{rc sig}
 \begin{split}
  \dv{\Tilde{\sigma}_{\mu\nu}}{\tau} =  u^\rho\nabla_\rho\tilde{\sigma}_{\mu\nu}=-\frac{2}{3} \tilde{\theta}\tilde{\sigma}_{\mu\nu}-\tilde{\sigma}_{\mu\rho}{\tilde{\sigma}^\rho}_\nu-\tilde{\omega}_{\mu\rho}{\tilde{\omega}^\rho}_\nu &+\frac{1}{3}h_{\mu\nu}\left(\tilde{\sigma}_{\rho\lambda}\tilde{\sigma}^{\rho\lambda}-\tilde{\omega}_{\rho\lambda}\tilde{\omega}^{\rho\lambda}\right)\qquad \\
&\qquad +\tilde{C}_{\rho\nu\mu\lambda}u^\rho u^\lambda+\frac{1}{2}\tilde{\mathcal{R}}_{\mu\nu},
 \end{split}
\end{equation}
where $\tilde{\mathcal{R}}_{\mu\nu}$ is the trace-free, spatial part of $\tilde{R}_{\mu\nu}$,
\begin{equation}
 \tilde{\mathcal{R}}_{\mu\nu}=h_{\mu\rho}h_{\nu\lambda}\tilde{R}^{\rho\lambda}-\frac{1}{3}h_{\mu\nu}h_{\rho\lambda}\tilde{R}^{\rho\lambda},
\end{equation}
and $C_{\mu\nu\rho\lambda}$ is the {\it Weyl tensor}, defined as
\begin{equation}
 \tilde{C}_{\mu\nu\rho\lambda}=\tilde{R}_{\mu\nu\rho\lambda}+\frac{1}{2}\left(\tilde{R}_{\mu\lambda}g_{\nu\rho}-\tilde{R}_{\mu\rho}g_{\nu\lambda}+\tilde{R}_{\nu\rho}g_{\mu\lambda}-\tilde{R}_{\nu\lambda}g_{\mu\rho}\right)+\frac{1}{6}\tilde{R}\left(g_{\mu\rho}g_{\nu\lambda}-g_{\mu\lambda}g_{\nu\rho}\right).
\end{equation}
The evolution equation for rotation is
\begin{equation}\label{rc om}
 \dv{\Tilde{\omega}_{\mu\nu}}{\tau} = u^\rho\tilde{\nabla}_\rho\tilde{\omega}_{\mu\nu}=-\frac{2}{3}\tilde{\theta} \tilde{\omega}_{\mu\nu}-2{\tilde{\sigma}^\rho}_{[\nu}\tilde{\omega}_{\mu]\rho}.
\end{equation}
Sometimes the term Raychaudhuri equations is used to refer to all of the three equations \eqref{rceq}, \eqref{rc sig} and 
\eqref{rc om}.

For hypersurface orthogonal geodesics, we have $\tilde{\omega}_{\mu \nu} = 0$. We consider the theory of gravity to be General Relativity. When the matter distribution satisfies the strong energy condition  $T_{\mu\nu}u^\mu u^\nu+\frac{1}{2}T\geq 0$, we have $\tilde{R}_{\mu\nu} u^\mu u^\nu \geq 0$ by virtue of the Einstein equations. Consequently, the right hand side of (\ref{rceq}) is negative because $\tilde\sigma_{\mu\nu}\tilde\sigma^{\mu\nu}\geq 0$ for spatial $\tilde\sigma_{\mu\nu}$\,. It follows that if the congruence is initially
converging, i.e. $\tilde{\theta}_0 \equiv \tilde{\theta}(0) < 0$, the geodesics will focus  within a finite value of the proper time $\tau \leq {3}{\tilde{\theta}_0}^{-1}$. This concept is popularly known as the focusing theorem~\cite{Poisson:2009pwt, Wald:1984rg} and the focusing condition is mathematically written as
\begin{equation}
 \frac{ \mathrm{d}{\Tilde{\theta}}}{{\mathrm{d}\tau} }+ \frac{1}{3} \Tilde{\theta}^2\leq 0.
\end{equation}
Focusing of a congruence implies existence of singularities in a spacetime which satisfy a few other physically reasonable conditions~\cite{Penrose:1964wq, Hawking:1970zqf, hawking_ellis_1973}.

In a gravitational collapse model, we often choose a congruence of the worldlines of the collapsing particles inside the matter distribution~\cite{Choudhury:2019snb, Choudhury:2022uek}. If the congruence focuses within a finite time then occurrence of a singularity is inevitable. Violation of the focusing condition, on the other hand,
comes with a possibility of avoiding the singularity formation. 
A crucial assumption behind the focusing theorem is the convergence condition $\tilde{R}_{\mu\nu} u^\mu u^\nu \geq 0$ and again this is implied from the strong energy condition. When the effective interaction is incorporated in the energy-momentum tensor, there is a possibility of violation of the strong energy condition as the collapse proceeds and density grows. This is because the interaction is expected to become effective in these situations. If the strong energy condition is violated, the convergence condition no more holds and focusing can be prevented. We, therefore, investigate the fate of the strong energy condition in this case. We split the energy-momentum tensor for fermionic fields given in Eq.~\eqref{emt} as
\begin{equation}
 T_{\mu\nu}=T^\mathrm{free}_{\mu\nu}+T^\mathrm{int}_{\mu\nu},
\end{equation}
where
\begin{equation}\label{freeEMT}
 T^\mathrm{free}_{\mu\nu} 
 { = \frac{1}{4} \sum_{f}^{}\left[\Bar{\psi}_f \gamma_{(\mu} \Tilde{D}_{\nu)} \psi_f + \text{h.c.}\right]\,,}
\end{equation}
and
\begin{equation}\label{intT}
 { T^\mathrm{int}_{\mu\nu} =\frac{1}{2}g_{\mu\nu}\left[\sum_f\left(\lambda_{fV} \Bar{\psi}_f\gamma_I\psi_f+\lambda_{fA} \Bar{\psi}_f\gamma_I\gamma_5\psi_f\right)\right]^2.}
\end{equation}
We can expect that the free part $T^\mathrm{free}_{\mu\nu}$\,, being the energy momentum tensor of a gas of free fermions, satisfies the strong energy condition. So this part will provide an attracting contribution in the Raychaudhuri equation and will help in focusing. We shall verify this demand later in this paper. For the interaction part we have
\begin{equation}
  { \rho^\mathrm{int}\equiv T^\mathrm{int}_{\mu\nu}u^\mu u^\nu = -\frac{1}{2}\left[\sum_f\left(\lambda_{fV} \Bar{\psi}_f\gamma_I\psi_f+\lambda_{fA} \Bar{\psi}_f\gamma_I\gamma_5\psi_f\right)\right]^2.}
\end{equation}
{We note that despite appearances, this is not negative definite, as the axial current can be spacelike.} {This also depends on the $\lambda_{fV}$, $\lambda_{fA}$ can be positive or negative.}
It is known that $\rho\equiv T_{\mu\nu}u^\mu u^\nu$ is the energy density seen by an observer moving with velocity $u^\mu$. If the interaction part  provides a non-negative contribution to the effective total energy density of the fermions, we should have $T^\mathrm{int}_{\mu\nu} u^\mu u^\nu\geq 0$. This means that the contribution of the interaction part to the effective energy-momentum tensor should satisfy the weak energy condition. Let us now examine the strong energy condition for the interaction part. It follows from Eq.~\eqref{intT} that
\begin{equation}\label{SECint}
 T^\mathrm{int}_{\mu\nu} u^\mu u^\nu+\frac{1}{2}T^\mathrm{int}=-T^\mathrm{int}_{\mu\nu} u^\mu u^\nu.
\end{equation}
Consequently, when the contribution from the interaction part satisfies the weak energy condition, this contribution must violate the strong energy condition. {Looking at the form of $T^\mathrm{int}_{\mu\nu}$ we can write}
\begin{equation}
 p^\mathrm{int}=-\rho^\mathrm{int}.
\end{equation}
%
%
If $\rho^\mathrm{int}\geq 0$, we can interpret that the effective equation of state for the interaction part behaves like that of a~{positive}
Cosmological Constant and provide an effective negative pressure. This fact is discussed by Poplawski in \cite{Poplawski:2010jv, Poplawski:2011wj}. 
Therefore, it will provide a repulsive contribution in the Raychaudhuri equation. This can help to prevent the focusing of the congruence. 
This inference is true for any geometry as we have not assumed any specific kind of spacetime until now. As the free part helps in focusing, the 
ultimate fate of the collapse will then depend on the competition between the free and interaction parts. If a situation appears such that the 
interaction part dominates, the focusing can, in principle, be prevented. However, at first we should check if the interaction part satisfies the 
weak energy condition. For this, we shall assume a distribution containing different species of fermions which is collapsing under gravity. We calculate the expectation values of
the free and interaction part of the energy-momentum tensor with respect to $n$-particle states. We are dealing with classical physics in the
geometry part. Hence if we intend to replace the geometry part ($R_{\mu\nu}u^\mu u^\nu$) in the Raychaudhuri equation by the energy-momentum 
tensor of the fermionic field via Einstein's equations, we should take the relevant expectation values.

 {\section{Energy conditions and Fate of the focusing condition}\label{sec4}}
To get insights about the energy conditions, at first we need to calculate the expectation value of the energy momentum tensor $\left\langle T_{\mu\nu}\right\rangle$ with respect to a suitably chosen state describing the system { of a gas of fermions}. 
For the interaction part this is given by
\begin{equation}\label{EMT1}
  \left\langle T^\mathrm{int}_{\mu\nu}\right\rangle = \frac{1}{2} g_{\mu\nu}\sum_{f,f'} \Big[ \lambda_{fV} \lambda_{f'V}  \left\langle J^{I}_f J_{f'I}\right\rangle + \lambda_{fA}\lambda_{f'A} \left\langle J^{I}_{f5} J_{f'I5}\right\rangle + 2\lambda_{fV} \lambda_{f'A} \left\langle J^{I}_f J_{f'I5}\right\rangle 
  \Big],
\end{equation}
where, $J_{fI} \equiv \Bar{\psi}_f \gamma_I \psi_f$ and $J_{fI5} \equiv \Bar{\psi}_f \gamma_I \gamma_5 \psi_f$.
{ We are not aware of any method to exactly calculate} this expectation value with respect to a sufficiently 
general state describing multiple species of fermions.
{ Assuming the interaction strength is small, currents of different species may be treated as independent, so for  $f\neq f'$  we can write }
\begin{equation}\label{expass1}
 \left\langle J^{I}_f J_{f'I}\right\rangle=\left\langle J^{I}_f \right\rangle \left\langle  J_{f'I}\right\rangle, \hspace{0.2cm}  \left\langle J^{I}_{f5} J_{f'I5}\right\rangle=\left\langle J^{I}_{f5} \right\rangle \left\langle  J_{f'I5}\right\rangle, \hspace{0.2cm}  \left\langle J^{I}_f J_{f'I5}\right\rangle=\left\langle J^{I}_f \right\rangle \left\langle  J_{f'I5}\right\rangle.
\end{equation}
However, the above decomposition cannot be done  for the same species, i.e. when $f=f'$\,. {In special cases, a condensate might form, 
based on pairing between states of opposing spins as in the BCS theory of superconductivity or between particles and antiparticles as in the 
NJL model of mesons. Then the effective interaction has scalars and vectors coupled to fermion bilinears. On the other hand, the interaction we are 
considering is expected to be quite weak, so pair formation is unlikely. Therefore, we explore a different approach to calculating
the contribution of fermion self-interactions to $\langle T^\mathrm{int}_{\mu\nu}\rangle$\,. }


We will calculate this for one species of fermion and then obtain $ \left\langle T^\mathrm{int}_{\mu\nu}\right\rangle$ by carrying out the sum in Eq.~\eqref{EMT1}. For convenience, we will drop the index specifying different species in the following where we calculate the relevant expectation values for a single species. We will restore this index when we talk about multiple species.

We will calculate the expectation values by considering the Fock space quantization formulation.
Our primary aim is to look into the nature of the interaction term, specifically whether it gives a repulsive or attractive contribution.
We need to consider a simplified scenario for this purpose, {since solutions of the Dirac equation $\psi$ in collapsing spacetimes, 
even without fermion interactions, are hardly discussed in the literature.  
 {We assume the density of fermions to be small enough and the
	geometrical interaction to be weak enough  that we can expand the fermion fields in terms of plane waves on flat Minkowski space.} 
We will suppress all other interactions of the fermions.
These assumptions are reasonable at least at the { early}
stages of collapse and capable of revealing the
 qualitative nature of the free and interaction terms. Hence we write the Dirac field and its conjugate as
\begin{equation}\label{solpsi}
\begin{split}
 &\psi (x) = \sum_s \int d^3k \sqrt{\frac{1}{(2\pi)^3 V k^0}} \Big[ e^{ i k\cdot x} c(\bm{k}, s) u(k, s) + e^{-i k\cdot x}  d^\dagger(\bm{k}, s) v(k, s) \Big],\\
 &\Bar{\psi} (x) = \sum_{s'} \int d^3k' \sqrt{\frac{1}{(2\pi)^3 V {k'}^0}} \Big[ e^{-i k'\cdot x} c^\dagger(\bm{k'}, s') \Bar{u}(k', s') + e^{ i k'\cdot x} d(\bm{k'}, s') \Bar{v}(k', s') \Big],
\end{split}
\end{equation}
where $k^0=\sqrt{k^2+m^2}$ and $V$ is the volume of the distribution. 
{Note that a four-fermion interaction does not lead to a consistent quantum field theory, i.e. one
which is valid at all energy scales,  in a flat spacetime. We will assume that the energies involved in our calculations
are small enough that we can restrict to tree-level results.
	We will come back to this point in the last section. }
Let us consider a state with $n$ fermions, given by
\begin{equation}\label{npstate}
 \ket{n} = \prod_{i=1}^n c^\dagger(\bm{k_i}, s_i) \ket{0}\,.
\end{equation}
If we consider a gas where both spins of the fermions are equally probable and the anti-fermions are absent,
it follows from a lengthy but straightforward calculation  
(see Appendix~\ref{appc} for details) that
%
\begin{align}\label{expectation2}
 \left\langle J^{I}J_{I}\right\rangle\equiv   \left\langle n \left| J^{I}J_{I}\right| n\right\rangle= &\sum\limits_{j=1}^n \int \frac{d^3k}{V} \frac{m^2}{(2\pi)^6} \frac{1}{k^0 k_j^0} \Big( -2 - \frac{1}{m^2} k^I k_{jI} \Big)  \notag\\ 
  &\qquad + \sum_{\substack{i,j=1\\ i \neq j}}^n \frac{m^2}{(2\pi)^6 V^2} \frac{1}{k_i^0 k_j^0} \left( 1 + \frac{3}{2m^2} k^{iI} k_{jI}\right).
\end{align}
{ There is an apparent divergence  in the above expression which can be resolved if we consider the gas to be at finite temperature. This is an important assumption in the following calculation which we will discuss shortly.}

A similar calculation for $J_{I5}$ yields (refer to Appendix~\ref{appd} for details)
\begin{equation}\label{ax5}
 {\left\langle J^{I}_5J_{I5}\right\rangle = \sum_{j=1}^n \int \frac{d^3k}{V} \frac{m^2}{(2\pi)^6} \frac{1}{k^0 k_j^0} \Big( 2 - \frac{1}{m^2} k^I k_{jI}\Big) - \frac{1}{2} \sum_{\substack{i, j=1 \\ i\neq j}}^n \frac{m^2}{(2\pi)^6 V^2} \frac{1}{k_i^0 k_j^0} \Big( 2 - \frac{1}{m^2} k_i^I k_{jI}\Big)\,.}
\end{equation}
Finally, for the cross term, we have {(see Appendix~\ref{appcross})}
 \begin{equation}\label{crossc}
{\left\langle J^{I5}J_{I}\right\rangle=0.}
\end{equation}    

We will assume that we have a large number of particles, so we can replace the discrete summations by integrations in the next step. Moreover, we will assume that the distribution is sufficiently hot such that we can incorporate the Maxwell-Boltzmann distribution function within the integration. Consequently, replacing $\sum_i \rightarrow V\int d^3k ~e^{-\beta k^0}$ we can write Eq.~(\ref{expectation2}) as
%
\begin{equation}\label{expectation3}
 {\left\langle J^{I}J_{I}\right\rangle = -\frac{m^2}{(2\pi)^6} \int d^3k' \int d^3k \frac{1}{k^0 k'^0} e^{-\beta k^0} e^{-\beta k'^0} + \frac{1}{2} \frac{1}{(2\pi)^6} \int d^3k' \int d^3k \frac{(k^I k'_I)}{k^0 k'^0} e^{-\beta k^0} e^{-\beta k'^0},}
\end{equation}
whereas Eq.~\eqref{ax5} yields, 
\begin{equation}\label{jiji5}
 {\left\langle J^{I5}J_{I5}\right\rangle = - \left\langle J^{I}J_{I}\right\rangle.}
\end{equation}
Here $\beta\sim \frac{1}{\mathcal{T}}$, where $\mathcal{T}$ denotes temperature of the matter distribution.

We can carry out the integrations in Eq. \eqref{expectation3} considering spherical polar coordinates:
\begin{equation}
 \begin{split}
  &\vec{k} = \big( k \sin \theta \cos \phi,\enspace  k \sin \theta \sin \phi, \enspace k \cos \theta \big),\\
  &\vec{k'} = \big( k' \sin \theta' \cos \phi',\enspace  k' \sin \theta' \sin \phi', \enspace k' \cos \theta' \big),
 \end{split}
\end{equation}
and $d^3k = k^2 \sin \theta dk d\theta d\phi$, $d^3k' = k'^2 \sin \theta' dk' d\theta' d\phi'$. This yields,
\begin{equation}\label{expectation4}
 \left\langle J^{I}J_{I}\right\rangle = -\frac{m^4}{4\pi^4\beta^2} \Big[ K_1(\beta m)\Big]^2 - \frac{m^4}{8\pi^4\beta^2} \Big[ K_2(\beta m) \Big]^2\equiv -A(\beta, m)-B(\beta, m),
\end{equation}
where $K_\nu(z)$ is the modified Bessel function of the second kind and we have defined $A(\beta, m)=\frac{m^4}{4\pi^4\beta^2} \Big[ K_1(\beta m)\Big]^2$ and $B(\beta, m)=\frac{m^4}{8\pi^4\beta^2} \Big[ K_2(\beta m) \Big]^2$. It is quite evident that we always have  $\left\langle J^{I}J_{I}\right\rangle\leq 0$. The plots of $A$ and $B$ with $\beta$ is shown in Fig. \ref{fig1}.
\begin{figure}[H]
\centering
\boxed{\includegraphics[width=0.7\textwidth]{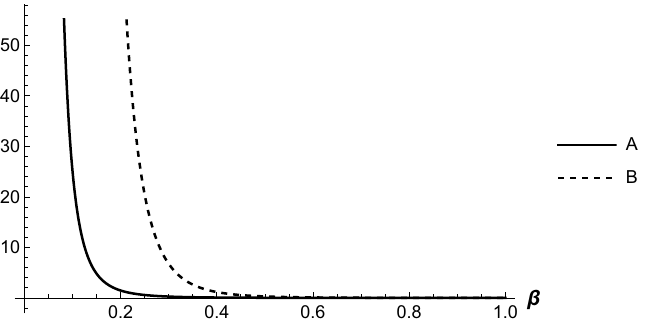}}
 \caption{Plot of $A$ and $B$ with $\beta$ for $m=1$.}
 \label{fig1}
\end{figure}

Now, for high temperatures (i.e. for small values of $\beta$), $K_1( \beta m) \approx \frac{1}{\beta m}$, whereas $K_2( \beta m) \approx \frac{1}{\beta^2 m^2}$. So, Eq.~\ref{expectation4} approximates as
\begin{equation}\label{expjiji}
 \left\langle J^{I}J_{I}\right\rangle \sim -(C_1 m^2 {\mathcal{T}^4} + C_2 {\mathcal{T}^6}),
\end{equation}
where $C_1, C_2>0$ but are the same for all fermions when $T\gg m\,.$

Similar calculation for $\left\langle J^I\right \rangle \equiv \left\langle n\left|J^I\right|n\right\rangle$ and $\left\langle J^I_5\right \rangle \equiv \left\langle n\left|J^I_5\right|n\right\rangle$ yields,
\begin{equation}\label{expjiji5}
 \left\langle J^I\right \rangle=\frac{1}{(2\pi)^3}\int \frac{d^3 k}{k^0} k^I e^{-\beta k^0}, \hspace{0.2cm} \left\langle J^I_5\right \rangle=0.
\end{equation}
At high temperatures,
\begin{equation}\label{expjihight}
  {\left\langle J^I\right \rangle\sim  \mathcal{T}^3}.
\end{equation}



Using all these results we get from Eq.~\eqref{EMT1} at large $\mathcal{T}$\,,
\begin{equation}\label{edint}
\begin{split}
 {\left\langle T^\mathrm{int}_{\mu\nu}\right\rangle u^\mu u^\nu \sim \sum_f \frac{1}{2} \big(  \lambda_{fV}^2-\lambda_{fA}^2 \big) (C_{1} m_f^2 {\mathcal{T}^4} + C_{2} {\mathcal{T}^6})+\sum_{\substack{f, f' \\ f\neq f'}}\lambda_{fV}\lambda_{f'V} C_{3} \mathcal{T}^6}
 \end{split}
\end{equation}
{where $C_3>0$}.
The terms proportional to $\mathcal{T}^6$ in the above expression gives the dominating contribution in the energy density coming from the interaction part $\left(\left\langle T^\mathrm{int}_{\mu\nu}\right\rangle  u^\mu u^\nu\right)$ at high temperature regime. Dolan has mentioned a similar temperature dependence of the energy density, due to a four-fermion interaction with a single species, in the context of FLRW cosmology~\cite{Dolan:2009ni}.  Now, from Eq. \eqref{SECint}, we write
\begin{equation}\label{secintf}
 { \left\langle T^\mathrm{int}_{\mu\nu}\right\rangle u^\mu u^\nu +\frac{1}{2}\left\langle T^\mathrm{int}\right\rangle\sim  \sum_f \frac{1}{2} \big(  \lambda_{fA}^2-\lambda_{fV}^2 \big) (C_{1} m_f^2 {\mathcal{T}^4} + C_{2} {\mathcal{T}^6})-\sum_{\substack{f, f' \\ f\neq f'}}\lambda_{fV}\lambda_{f'V} C_{3} \mathcal{T}^6},
\end{equation}
where $\left\langle T^\mathrm{int}\right\rangle=g^{\mu\nu}\left\langle T^\mathrm{int}_{\mu\nu}\right\rangle$. Eq. \eqref{edint} suggests that the energy density due to the interaction part can be positive or negative (and hence the contribution from interaction is repulsive or attractive respectively according to Eq.~\eqref{secintf})  depending on $\lambda_V$s and $\lambda_A$s of different species. Let us suppose that the geometrical coupling is maximally chiral, i.e., for each species, $\lambda_{fA}=\pm\lambda_{fV}$. Then for a distribution of fermions, we find 
%
%
\begin{equation}\label{edint}
\begin{split}
 { \left\langle T^\mathrm{int}_{\mu\nu}\right\rangle u^\mu u^\nu=-\left[\left\langle T^\mathrm{int}_{\mu\nu}\right\rangle u^\mu u^\nu +\frac{1}{2}\left\langle T^\mathrm{int}\right\rangle\right] \sim \sum_{\substack{f, f' \\ f\neq f'}}\lambda_{fV}\lambda_{f'V} C_{3} \mathcal{T}^6}.\end{split}
\end{equation}

We will now come to the free part of the energy-momentum tensor. Following  exactly the same procedure as for the interaction part, we have
\begin{equation}\label{freeexp}
\left\langle T_{\mu\nu}^\mathrm{free}\right\rangle=\sum_f\frac{2}{(2\pi)^3}\int d^3k_f~ e^{-\beta k^0_f}~\frac{k_{f\mu} k_{f\nu}}{k^0_f}.
\end{equation}
{The above expression for $\left\langle T_{\mu\nu}^\mathrm{free}\right\rangle$ matches with the results of~\cite{PhysRevD.108.044024, MARTIN2012566} at the zero temperature limit.}
Thus,
\begin{equation}
 \left \langle T_{\mu\nu}^\mathrm{free}\right \rangle u^\mu u^\nu= \sum_f\frac{2}{(2\pi)^3}\int d^3k_f~ e^{-\beta k^0_f}~\frac{\left(k_{f\mu} u^\mu\right)^2}{k^0_f},
\end{equation}
and
\begin{equation}\label{secfree}
 \left \langle T_{\mu\nu}^\mathrm{free}\right \rangle u^\mu u^\nu+\frac{1}{2}\left \langle T^\mathrm{free}\right \rangle=\sum_f \Bigg(\frac{2}{(2\pi)^3}\int d^3k_f~ e^{-\beta k^0_f}~\frac{\left(k_{f\mu} u^\mu\right)^2}{k^0_f}+\frac{1}{(2\pi)^3}\int d^3k_f~ e^{-\beta k^0_f}~\frac{\left(k_{f\mu} k^{\mu}_f\right)}{k^0_f}\Bigg).
\end{equation}
It can be shown from Eq.~\eqref{secfree} that at high temperatures,
\begin{equation}
{ \left \langle T_{\mu\nu}^\mathrm{free}\right \rangle u^\mu u^\nu+\frac{1}{2}\left \langle T^\mathrm{free}\right \rangle \sim  \mathcal{T}^4},
\end{equation}
implying that the free part satisfies the strong energy condition.\\

Let us now come back to the Raychaudhuri equation and the focusing condition in this case. From the Raychaudhuri equation \eqref{rceq} we have,
\begin{equation}
  \frac{ \mathrm{d}{\Tilde{\theta}}}{{\mathrm{d}\tau} }+ \frac{1}{3} \Tilde{\theta}^2=  - \tilde{\sigma}_{\mu\nu}\tilde{\sigma}^{\mu\nu}
 +\tilde{\omega}_{\mu\nu}\tilde{\omega}^{\mu\nu} - \Tilde{{R}}_{\mu\nu} u^{\mu} u^{\nu}.
\end{equation}
As we are assuming a nearly flat geometry, we neglect the effects due to shear and rotation. Also, it is reasonable to assume that the curvature driven effects will be dominant as the collapse proceeds. Therefore, using the Einstein's equations,
\begin{equation}
 \Tilde{{R}}_{\mu\nu}-\frac{1}{2}g_{\mu\nu}\Tilde{R}=\left\langle T_{\mu\nu}\right\rangle,
\end{equation}
we get
\begin{equation}
  \frac{ \mathrm{d}{\Tilde{\theta}}}{{\mathrm{d}\tau} }+ \frac{1}{3} \Tilde{\theta}^2\approx  - \left[\left \langle T_{\mu\nu}^\mathrm{free}\right \rangle u^\mu u^\nu+\frac{1}{2}\left \langle T^\mathrm{free}\right \rangle\right]-\left[\left\langle T^\mathrm{int}_{\mu\nu}\right\rangle u^\mu u^\nu +\frac{1}{2}\left\langle T^\mathrm{int}\right\rangle\right].
\end{equation}
Therefore, at high temperatures we find
\begin{equation}\label{fcf}
 { \frac{ \mathrm{d}{\Tilde{\theta}}}{{\mathrm{d}\tau} }+ \frac{1}{3} \Tilde{\theta}^2\sim {\sum_f \frac{1}{2} \big(  \lambda_{fV}^2-\lambda_{fA}^2 \big) (C_{1} m_f^2 {\mathcal{T}^4} + C_{2} {\mathcal{T}^6})+\sum_{\substack{f, f' \\ f\neq f'}}\lambda_{fV}\lambda_{f'V} C_{3} \mathcal{T}^6}-C_4 \mathcal{T}^4 \,,}
\end{equation}
{where all of the constants $C_1$, $C_2$, $C_3$ and $C_4$ are positive}.

At high temperatures, we can expect the interaction part to provide the dominant contribution through its $\mathcal{T}^6$ dependence, then the focusing condition is violated when $\lambda_V$'s and $\lambda_A$'s are such that  $\left\langle T^\mathrm{int}_{\mu\nu}\right\rangle u^\mu u^\nu +\frac{1}{2}\left\langle T^\mathrm{int}\right\rangle<0$. {For example, when the distribution is maximally chiral, the focusing condition will be violated when the quantity $\sum_{\substack{f, f' \\ f\neq f'}}\lambda_{fV}\lambda_{f'V}>0$. Or if there is only one species of fermions in the distribution, the focusing condition can be violated when $\lambda_{A}^2<\lambda_{V}^2$  for that particular species.} Consequently, a final singularity may be prevented  in these cases.
However, if the distribution carries only one species of fermions, with maximally chiral geometrical coupling, the contribution from the interaction term vanishes and formation of a singularity is inevitable. {It is worthwhile to mention that for the case of purely axial current (i.e., $\lambda_{V}=0$), the contribution from the four-fermion interaction is attractive which is evident from Eq. \eqref{fcf}. This matches with the findings of \cite{Kerlick:1975tr, PhysRevD.16.1247} which deals with purely axial currents. The possibility of a repulsive contribution comes due to the presence of a mixed current which arises from the assumption that contorsion can couple with different species and chiralities with diferent strengths.}

At this stage, we must emphasize that we have assumed that the fermions are in a dilute thermal distribution. 
As the collapse proceeds, the density increases and the assumption of the nearly flat spacetime and small effects of the interaction 
term may not hold. In spite of that, we can intuitively argue that  the interaction term can give rise to repulsive effects depending 
on the coupling constants ($\lambda_{fV}$, $\lambda_{fA}$) involved. {For stars with large enough densities,} this interaction term is expected to dominate at high temperatures, and may provide a means to avoid the final singularity formation. Nevertheless, we need to consider the assumption of curved spacetime and non-negligible contribution from the interaction term for a definite conclusion in this regard. This is quite difficult as well as complicated and out of the purview of the current work.\\

\section{Conclusion and final remarks}\label{sec5}
In this work, we have examined the effect of dynamically generated torsion, which arises due to presence of fermions in curved spacetime, on the gravitational collapse of a fermionic distribution. Specifically, we have discussed if this torsion can give rise to repulsive effects and prevent the final singularity formation. For this we have used the approach of separating the torsion degree of freedom and replacing it via an effective four-fermion interaction in the energy-momentum sector while keeping the geometry part Riemannian. Therefore, we have carried out our work using the field equations of General Relativity with a modified energy-momentum tensor of the fermionic field.
For getting insights about the gravitational collapse, we have adopted the approach of using the Raychaudhuri equation and the focusing condition.

Focusing of a congruence, which is essential in proving the existence of singularities, assumes that the strong energy condition is obeyed. For the geometrical  interaction we have considered,  the modified energy-momentum tensor carries an effective interaction term along with the free fermionic part. We have examined what are the characteristics of this interaction part, specifically if it can violate the strong energy condition and dominate at higher densities. For this, we have chosen a dilute distribution of large number of fermions collapsing under gravity. The dynamically generated torsion
couples to the different species, and chiralities, of the fermions with different strengths. We have assumed that the distribution is sufficiently hot
that it obeys the Maxwell-Boltzmann distribution and all the spins are equally probable. In addition, we have considered the distribution to be large and dilute enough so that we can continue with the assumption of a nearly flat geometry with small effects due to the interaction term. The final assumption is valid at the early  stages of collapse as well as for  a significant time as the collapse proceeds. It enables us to use the well-known flat spacetime solution for the fermionic field $\psi$ and calculate the necessary expectation values needed to get insights about the energy conditions. This is worthwhile as we can extract essential information about the system under consideration when exact solutions for $\psi$ for such systems are hardly known. We can also naively discuss about the limits until which the assumption of a nearly flat geometry holds. For instance, let us argue that this flatness holds upto a limit when the metric coefficients ($g_{tt}$ or $g_{rr}$, assuming a Schwarzschild like geometry) differ from unity at most by an order of $10^{-3}$. Then the corresponding mass to radius limit of the relevant distribution comes out to be of the order of $10^{24}$ kg/m$^3$. White dwarfs falls well below this limit, whereas for neutron stars this assumption of flatness will not be valid.

{Our use of quantum field theory in the calculation also calls for some comments. A four-fermion  interaction is clearly nonrenormalizable. However, while that is an accurate statement in flat spacetime, it may not be applicable in the case we have considered. The reason is that this interaction arises from the dynamics of fermions  in curved spacetime; it would not have appeared if we had started with a flat background. It appears because torsion,  which may be thought  of as one of the gauge fields of local Poincar\'e symmetry, is a nondynamical field.  Moreover, a renormalizable theory is one which remains structurally same at all length scales, up to redefinitions of couplings and masses. Nonrenormalizability of this interaction will be important at high energies, and because it originates from the fact that spacetime is curved, we can expect it to be resolved analogously to the similar problem of quantum gravity, e.g.  by having a fixed point, as in the asymptotic safety conjecture of Weinberg~\cite{Weinberg:1980gg, Niedermaier:2006wt}. However, we have considered dilute gases at low energy so that loop effects can be ignored.}

The major conclusions of this examination are as follows. Firstly, if the energy density due to the interaction part is positive, i.e., it satisfies the weak energy condition, the strong energy condition must be violated. In that case, the interaction term will behave analogous to an exotic matter component, having an effective negative equation of state.  This conclusion actually depends on how contorsion couples with different chiralities of the fermions present in the distribution, i.e., on the sign of the coupling constants, $\lambda_{fV}$ and $\lambda_{fA}$. The interaction part contributes an effective negative pressure which acts against the collapse if the strong energy condition is violated. Another important conclusion which we have established is that the interaction term starts to dominate as the collapse proceeds and at high temperatures, it provides the dominating contribution. However, for the special case when there exists only one species of maximally chiral fermions, the contribution from the interaction term vanishes. Therefore, it is, in principle, possible to halt the collapse and prevent the final singularity formation when the coupling constants $\lambda_{fV}$ and $\lambda_{fA}$ are such that the interaction part violates the strong energy condition. 


We conclude by mentioning the following essential point. At the final stages of collapse, the curvature and the densities should be high and our assumption of a nearly flat geometry with a small effect from the interaction part will not hold. Also, there may be non-vanishing contributions from the shear and rotation parts which can play a significant role as the collapse proceeds. Therefore, for a concrete establishment of the conclusion that the effective four-fermion interaction provides a repulsive contribution and can halt a collapse, we must work in curved geometries addressing all of the above facts along with considering the non-linear Dirac equation \eqref{ndceq}. However, the current work acquires importance since it allows us to draw
important conclusions about fermionic collapsing systems which incorporates effective four-fermion interaction, in spite of the fact that no exact
solution for such systems seems to be available in the literature. Also, the conclusions are expected to be valid for a significant time during the collapse of realistic stars.

\section*{Acknowledgments}
The authors thank Subhasish Chakrabarty and Shantonu Mukherjee for useful discussions regarding this work. SGC is also thankful to Ananda Dasgupta for providing
valuable insights.
\appendix
\renewcommand{\theequation}{\thesection.\arabic{equation}}
%
\section{Derivation of the effective Energy-Momentum tensor}\label{appb}
The matter part of the action including the interaction term is given by,
\begin{equation}
\begin{split}
 S_\mathrm{m}=\int|e| \mathrm{d}^4 x &\left[ -\frac{1}{2}\left(\Bar{\psi} \gamma^K e^\mu_K (\Tilde{D}_\mu \psi) + (\Bar{\psi} \gamma^K e^\mu_K \Tilde{D}_\mu \psi)^\dagger +2 m \bar{\psi} \psi \right)\right.
    \\ & ~~\left.  -\frac{1}{2}\left(\lambda_V \Bar{\psi}\gamma_I\psi+\lambda_A \Bar{\psi}\gamma_I\gamma_5\psi\right)\left(\lambda_V\Bar{\psi}\gamma^I\psi+\lambda_A\Bar{\psi}\gamma^I\gamma_5\psi\right)\right]
     \end{split}
\end{equation}
If we vary this action with respect to tetrad $e^\mu_K$, we find
\begin{equation}\label{varyac}
 \delta S=\int \mathrm{d}^4 x \left[ \left|e\right| \left({X_\mu}^K \delta e^\mu_K+\frac{1}{2}\delta {\omega_{\mu}}^{IJ} {S^K}_{IJ}e^\mu_K \right) + \delta |e| \mathcal{L}_\mathrm{m} \right],
\end{equation}
where 
\begin{equation}\label{Xeq}
 {X_\mu}^K=-\frac{1}{2}\left(\Bar{\psi} \gamma^K  (\Tilde{D}_\mu \psi)  +  (\Bar{\psi} \gamma^K  \Tilde{D}_\mu \psi)^\dagger\right),
 \end{equation}
\begin{equation}\label{cont}
 {S^K}_{IJ}=-\frac{1}{8}\bar{\psi}\left[\gamma^K,\left[\gamma_I, \gamma_J\right]_-\right]_+\psi,
\end{equation}
and 
\begin{equation}
\begin{split}
 \mathcal{L}_\mathrm{m}=&-\frac{1}{2}\left(\Bar{\psi} \gamma^K e^\mu_K (\Tilde{D}_\mu \psi) + (\Bar{\psi} \gamma^K e^\mu_K \Tilde{D}_\mu \psi)^\dagger +2 m \bar{\psi} \psi \right)
    \\ &-\frac{1}{2}\left(\lambda_V \Bar{\psi}\gamma_I\psi+\lambda_A \Bar{\psi}\gamma_I\gamma_5\psi\right)\left(\lambda_V\Bar{\psi}\gamma^I\psi+\lambda_A\Bar{\psi}\gamma^I\gamma_5\psi\right).
     \end{split}
\end{equation}
Here symmetric and anti-symmetric brackets are denoted using curly and square brackets, respectively, in the expression \eqref{cont}.

The spin connection ${\omega_\mu}^{IJ}$ can be written in terms of the vierbein as,
\begin{equation}
 {\omega_\mu}^{IJ}=\frac{1}{2}e^{\nu I}\left(\partial_\mu e_\nu^J-\partial_\nu e_\mu^J\right)-\frac{1}{2}e^{\nu J}\left(\partial_\mu e_\nu^I-\partial_\nu e_\mu^I\right)-\frac{1}{2}e^{\rho I}e^{\sigma J}\left(\partial_\rho e_{\sigma K}-\partial_\sigma e_{\rho K}\right)e_\mu^K,
\end{equation}
which implies
\begin{equation}\label{spccon}
 {\omega_\mu}^{IJ}e^\mu_K=\frac{1}{2}e^{\nu I}e^\mu_K\left(\partial_\mu e_\nu^J-\partial_\nu e_\mu^J\right)-\frac{1}{2}e^{\nu J}e^\mu_K\left(\partial_\mu e_\nu^I-\partial_\nu e_\mu^I\right)-\frac{1}{2}e^{\mu I}e^{\nu J}\left(\partial_\mu e_{\nu K}-\partial_\nu e_{\mu K}\right).
\end{equation}
 We denote the first, second and third term in the right hand side of \eqref{spccon} as $P/2$, $Q/2$ and $R/2$, respectively, for convenience.
Thus,
\begin{equation}
 \delta{\omega_\mu}^{IJ}e^\mu_K+{\omega_\mu}^{IJ}\delta e^\mu_K=\frac{1}{2}\left(\delta P-\delta Q-\delta R\right).
\end{equation}
It is easy to derive that
\begin{equation}
 \delta P= e^{\nu I}e^\mu_K\left(\tilde{D}_\mu \delta e_\nu^J-\tilde{D}_\nu \delta e_\mu^J\right)+\delta e^{\nu I}{{\omega_\nu}^J}_K-\delta e^\mu_K {\omega_\mu}^{JI}.
 \end{equation}
We will get similar expressions for $\delta Q$ and $\delta R$.
After a straightforward manipulation we have,
\begin{equation}
\begin{split}
 \delta{\omega_\mu}^{IJ}e^\mu_K=\frac{1}{2}e^{\nu I}e^\mu_K\left(\tilde{D}_\mu \delta e_\nu^J-\tilde{D}_\nu \delta e_\mu^J\right)-\frac{1}{2}e^{\nu J}e^\mu_K\left(\tilde{D}_\mu \delta e_\nu^I-\tilde{D}_\nu \delta e_\mu^I\right)\\ -\frac{1}{2}e^{\nu J}e^{\mu I}\left(\tilde{D}_\mu \delta e_{\nu K}-\tilde{D}_\nu \delta e_{\mu K}\right).
 \end{split}
\end{equation}
We can now put the above expression in the second term of the equation \eqref{varyac}, namely,
\begin{equation}
 \frac{1}{2}\int \mathrm{d}^4 x \left|e\right|\delta {\omega_{\mu}}^{IJ} e^\mu_K {S^K}_{IJ}.
\end{equation}
After that, if we carry out an integration by parts with the condition that the variation of the vierbein vanishes at the boundaries, we have,
\begin{equation}
 \int \mathrm{d}^4 x \left|e\right|\delta {\omega_{\mu}}^{IJ} e^\mu_K {S^K}_{IJ}=\int \mathrm{d}^4 x \left|e\right| e_\mu^I e^{\nu J}\tilde{D}_\nu{S^K}_{IJ}\delta e^\mu_K,
\end{equation}
where we have used the completely anti-symmetric property of ${S^K}_{IJ}$. Thus we have,
\begin{equation}
 \delta S=\int \mathrm{d}^4 x \left[ \left|e\right| \left({X_\mu}^K \delta e^\mu_K+\frac{1}{2}\delta {\omega_{\mu}}^{IJ} {S^K}_{IJ}e^\mu_K \right) + \delta |e| \mathcal{L}_\mathrm{m} \right]=-\int \mathrm{d}^4 x \left|e\right| {T_\mu}^K \delta e^\mu_K,
\end{equation}
where 
\begin{equation}
 {T_\mu}^K=-{X_\mu}^K-\frac{1}{2} e_\mu^I e^{\nu J}\tilde{D}_\nu{S^K}_{IJ}+e^K_\mu  \mathcal{L}_\mathrm{m}.
\end{equation}
So,
\begin{equation}\label{TIK}
 T_{IK}=e^\mu_I T_{\mu K}=-e^\mu_I X_{\mu K}-\frac{1}{2}e^\nu_J \tilde{D}_\nu {S^J}_{KI}+\eta_{IK}\mathcal{L}_\mathrm{m}.
\end{equation}
Now, using the identity,
\begin{equation}
 \left[\gamma_J,\left[\gamma_K,\gamma_J\right]_-\right]_-=4\left(\eta_{JK}\gamma_I-\eta_{JI}\gamma_K\right),
\end{equation}
and then putting the Dirac equation \eqref{ndceq} (and its adjoint) one can show that,
\begin{equation}
 e^\nu_J \tilde{D}_\nu {S^J}_{KI}=e^\nu_K X_{\nu I}-e^\nu_I X_{\nu K},
\end{equation}
where $X_{\mu I}$ can be obtained from equation \eqref{Xeq} by lowering the Lorentz index. Again, if we put Dirac equation \eqref{ndceq} then the last term in equation \eqref{TIK} becomes \begin{equation}
  \frac{1}{2}  \eta_{IK} \left(\lambda_V \Bar{\psi}\gamma_I\psi+\lambda_A \Bar{\psi}\gamma_I\gamma_5\psi\right)\left(\lambda_V\Bar{\psi}\gamma^I\psi+\lambda_A\Bar{\psi}\gamma^I\gamma_5\psi\right).                                    \end{equation}
Hence we get
\begin{equation}
\begin{split}
 T_{\mu\nu} &=e^I_\mu e^K_\nu T_{IK}\\ &=-\frac{1}{2}\left(e^K_\nu X_{\mu K}+e^K_\mu X_{\nu K}\right) +\frac{1}{2}  g_{\mu\nu} \left(\lambda_V \Bar{\psi}\gamma_I\psi+\lambda_A \Bar{\psi}\gamma_I\gamma_5\psi\right)\left(\lambda_V\Bar{\psi}\gamma^I\psi+\lambda_A\Bar{\psi}\gamma^I\gamma_5\psi\right).
 \end{split}
\end{equation}
This is the expression for the energy-momentum tensor for a single species of fermions and it leads to the final expression given in Eq. \eqref{emt} when we consider all the species.

\section{Calculation of $\left\langle J^I J_I\right\rangle$}\label{appc}
Using Eqs. \eqref{solpsi}, \eqref{npstate} and assuming a background distribution where anti-fermions are absent, we have,
\begin{equation}
\begin{split}
 J^I\ket{n}=\sum_{s,s^\prime} \int d^3k~  d^3k' \frac{1}{(2\pi)^3V\sqrt{k^0 {k^\prime}^0}}e^{-i(k^\prime-k).x} &\left[\Bar{u}(k', s')\gamma^I u(k,s)\right]\\
 &c^\dagger(\bm{k'}, s')c(\bm{k}, s)
 \prod_{i=1}^n c^\dagger(\bm{k_i}, s_i)\ket{0}.
\end{split}
\end{equation}
If we now repeatedly use the anti-commutation relation,
\begin{eqnarray}
	\left[ c(\bm{k},s), c^\dagger(\bm{k}^\prime,s^\prime)\right]_+ = \delta_{s s^\prime} ~\delta^3(\bm{k}-\bm{k}^\prime), 
\end{eqnarray}
with all other anti-commutators vanishing, we find
\begin{equation}\label{1stmat}
\begin{split}
 c^\dagger(\bm{k'}, s')c(\bm{k}, s) \prod_{i=1}^n c^\dagger(\bm{k_i}, s_i)\ket{0}=\sum_{j=1}^n\left(\prod_{i_1=1}^{j-1}c^\dagger(\bm{k_{i_1}}, s_{i_1})\right) c^\dagger(\bm{k'}, s')&\left(\prod_{i_2=j+1}^n c^\dagger(\bm{k_{i_2}}, s_{i_2})\right)\\&\delta_{s s_j} ~\delta^3(\bm{k}-\bm{k_j})\ket{0}.
 \end{split}
\end{equation}
Therefore, the final expression for $J^I$ is given by
\begin{equation}\label{jmun}
 J^I\ket{n}=\sum_{j=1}^n \sum_s\int d^3 k \frac{1}{(2\pi)^3V\sqrt{k^0 {k_j}^0}}e^{-i(k-k_j).x}\left[\Bar{u}(k, s)\gamma^I u(k_j,s_j)\right]{\prod_{l=1}^n}^\prime c^\dagger(\bm{k_l}, s_l)\ket{0},
\end{equation}
where ${\displaystyle \prod_{l=1}^n}^\prime$ signifies that in the product only the $j$-th term $c^\dagger(\bm{k_j}, s_j)$ will be replaced by $c^\dagger(\bm{k}, s)$.
Using Eq. \eqref{jmun} we can write,
\begin{equation}\label{expectation1}
 \left\langle J^{I}J_{I}\right\rangle = \bra{n} J^I J_I \ket{n} =- \sum_{j=1}^n \left(S_{jj} + \sum_{\substack{l=1\\ l \neq j}}^n S_{jl}\right),
\end{equation}
where,
\begin{equation}
\begin{split}
 S_{jj}\equiv \sum_{s,s^\prime}\int d^3k~ d^3k^\prime \Bigg[&\frac{ e^{i(k-k^\prime}).x}{(2\pi)^6 V^2 \sqrt{{k^\prime}^0 {k_j}^0}\sqrt{k^0 {k_j}^0}} \left[\Bar{u}(k, s)\gamma^I u(k_j,s_j)\right]\left[\Bar{u}(k_j, s_j)\gamma^I u(k^\prime,s^\prime)\right]\\  &\bra{0}\left(\prod_{i_1=n}^{j+1}c(\bm{k_{i_1}}, s_{i_1})\right) c(\bm{k'}, s') \left(\prod_{i_2=j-1}^{1}c(\bm{k_{i_2}}, s_{i_2})\right) \\ & \left(\prod_{i_3=1}^{j-1}c^\dagger(\bm{k_{i_3}}, s_{i_3})\right) c^\dagger(\bm{k}, s)\left(\prod_{i_4=j+1}^n c^\dagger(\bm{k_{i_4}}, s_{i_4})\right)\ket{0}\Bigg],
 \end{split}
\end{equation}
and
\begin{equation}
 \begin{split}
 S_{jl} (l\neq j)\equiv \sum_{s,s^\prime}\int d^3k~ d^3k^\prime \Bigg[&\frac{ e^{i(k-k_j).x}~ e^{-i(k'-k_l}).x}{(2\pi)^6V^2\sqrt{{k^\prime}^0 {k_l}^0}\sqrt{k^0 {k_j}^0}} \left[\Bar{u}(k, s)\gamma^I u(k_j,s_j)\right]\left[\Bar{u}(k_l, s_l)\gamma^I u(k^\prime,s^\prime)\right]\\ &\bra{0}\left(\prod_{i_1=n}^{j+1}c(\bm{k_{i_1}}, s_{i_1})\right) c(\bm{k'}, s') \left(\prod_{i_2=j-1}^{1}c(\bm{k_{i_2}}, s_{i_2})\right)\\ &\left(\prod_{i_3=1}^{l-1}c^\dagger(\bm{k_{i_3}}, s_{i_3})\right) c^\dagger(\bm{k}, s)\left(\prod_{i_4=l+1}^n c^\dagger(\bm{k_{i_4}}, s_{i_4})\right)\ket{0}\Bigg].
 \end{split}
\end{equation}
Now,
\begin{equation}
\begin{split}
 A_{jj}\equiv &\bra{0}\left(\prod_{i_1=n}^{j+1}c(\bm{k_{i_1}}, s_{i_1})\right) c(\bm{k'}, s') \left(\prod_{i_2=j-1}^{1}c(\bm{k_{i_2}}, s_{i_2})\right)\left(\prod_{i_3=1}^{j-1}c^\dagger(\bm{k_{i_3}}, s_{i_3})\right) c^\dagger(\bm{k}, s)\\ & \left(\prod_{i_4=j+1}^n c^\dagger(\bm{k_{i_4}}, s_{i_4})\right)\ket{0} =\delta_{s s^\prime} ~\delta^3(\bm{k}-\bm{k}^\prime)-\sum_{\substack{i=1\\ i \neq j}}^n \delta_{s s_i} ~\delta^3(\bm{k}-\bm{k}_i) \delta_{s^\prime s_i} ~\delta^3(\bm{k'}-\bm{k}_i),
 \end{split}
\end{equation}
and
\begin{equation}
 \begin{split}
 A_{jl}(l\neq j)= \bra{0}\left(\prod_{i_1=n}^{j+1}c(\bm{k_{i_1}}, s_{i_1})\right) c(\bm{k'}, s') \left(\prod_{i_2=j-1}^{1}c(\bm{k_{i_2}}, s_{i_2})\right)\left(\prod_{i_3=1}^{l-1}c^\dagger(\bm{k_{i_3}}, s_{i_3})\right) c^\dagger(\bm{k}, s)\\ \left(\prod_{i_4=l+1}^n c^\dagger(\bm{k_{i_4}}, s_{i_4})\right)\ket{0} =\delta_{s s_l} ~\delta^3(\bm{k}-\bm{k}_l) \delta_{s^\prime s_j} ~\delta^3(\bm{k'}-\bm{k}_j).
 \end{split}
\end{equation}
Using the above expressions for $A_{jj}$ and $A_{jl}$ in $S_{jj}$ and $S_{jl}$ respectively, we have after simplifications,
\begin{equation}
 \begin{split}
  S_{jj} = \sum_s \int d^3k & \frac{1}{(2\pi)^6 V k^0 k_j^0} \big[ \Bar{u}(k_j, s_j) \gamma^I u(k, s) \big] \big[ \Bar{u}(k, s) \gamma_I u(k_j, s_j)\big]\\ &- \sum_{\substack{i=1 \\ i \neq j}}^n \frac{1}{(2\pi)^6 V^2 k_i^0 k_j^0}  \big[ \Bar{u}(k_j, s_j) \gamma^I u(k_i, s_i) \big] \big[ \Bar{u}(k_i, s_i) \gamma_I u(k_j, s_j)\big],
  \end{split}
\end{equation}
\begin{equation}
  S_{jl} (l\neq j)= \frac{1}{(2\pi)^6 V^2 k_l^0 k_j^0} \big[ \Bar{u}(k_l, s_l) \gamma^I u(k_l, s_l) \big] \big[ \Bar{u}(k_j, s_j) \gamma_I u(k_j, s_j)\big].
\end{equation}
To get an idea about the nature and behaviour of the above terms, we consider the distribution to be thermal and containing a large number of fermions. Therefore, both the spins are equally probable and hence, at first, we will take spin averages of $S_{jj}$ and $S_{ij}$. The spin averages of these quantities are given by,
\begin{equation}
  \begin{split}
 {S}^\mathrm{SA}_{jj} \equiv \frac{1}{2}\sum_{s,s_j} \int d^3k &\frac{1}{(2\pi)^6 V k^0 k_j^0} \big[ \Bar{u}(k_j, s_j) \gamma^I u(k, s) \big] \big[ \Bar{u}(k, s) \gamma_I u(k_j, s_j)\big]\\ &-\frac{1}{4}\sum_{s_i, s_j} \sum_{\substack{i=1 \\ i \neq j}}^n  \frac{1}{(2\pi)^6 V^2 k_i^0 k_j^0} \big[ \Bar{u}(k_j, s_j) \gamma^I u(k_i, s_i) \big] \big[ \Bar{u}(k_i, s_i) \gamma_I u(k_j, s_j)\big],
  \end{split}
\end{equation}
and
\begin{equation}
  {S}^\mathrm{SA}_{jl} (l\neq j)= \frac{1}{4}\sum_{s_i, s_j} \frac{1}{(2\pi)^6 V^2 k_l^0 k_j^0} \big[ \Bar{u}(k_l, s_l) \gamma^I u(k_l, s_l) \big] \big[ \Bar{u}(k_j, s_j) \gamma_I u(k_j, s_j)\big].
\end{equation}
Using the following spin sum identities,
\begin{equation}
 \sum_s u(k, s) \Bar{u}(k, s) = \frac{\bm{-i\cancel{k}} + m}{2}, \quad \sum_s \Bar{u}(k, s) M u(k, s) = \Tr \Bigg[ M \frac{\bm{-i\cancel{k}} + m}{2} \Bigg],
\end{equation}
and identities involving gamma matrices, we have,
\begin{equation}
\begin{split}
 &\frac{1}{2} \sum_{s, s_j} \big[ \Bar{u}(k_j, s_j) \gamma^I u(k, s) \big] \big[ \Bar{u}(k, s) \gamma_I u(k_j, s_j)\big] =  \Big( 2m^2 + k^I k_{jI} \Big),\\
 &\frac{1}{4} \sum_{s_i, s_j} \big[ \Bar{u}(k_j, s_j) \gamma^I u(k_i, s_i) \big] \big[ \Bar{u}(k_i, s_i) \gamma_I u(k_j, s_j)\big] =  \frac{1}{2}\Big( 2m^2 +  k_i^{I} k_{jI} \Big),\\
 &\frac{1}{4} \sum_{s_i, s_j} \big[ \Bar{u}(k_i, s_i) \gamma^I u(k_i, s_i) \big] \big[ \Bar{u}(k_j, s_j) \gamma_I u(k_j, s_j)\big] =-  k_i^{I} k_{jI}.
\end{split}
\end{equation}
Therefore, the final expression for $\left\langle J^{I}J_{I}\right\rangle$ follows as,
\begin{equation}\label{expectationapp2}
{\begin{split}
 \left\langle J^{I}J_{I}\right\rangle=& -\sum_{j=1}^n \left({S}^\mathrm{SA}_{jj} + \sum_{\substack{l=1\\ l \neq j}}^n {S}^\mathrm{SA}_{jl}\right)\\ = &\sum_{j=1}^n \int d^3k \frac{m^2}{(2\pi)^6 V} \frac{1}{k^0 k_j^0} \Big( -2 - \frac{1}{m^2} k^I k_{jI} \Big) + \sum_{\substack{i,j=1\\ i \neq j}}^n \frac{m^2}{(2\pi)^6 V^2} \frac{1}{k_i^0 k_j^0} \\ &+ \frac{3}{2} \sum_{\substack{i,j=1\\ i \neq j}}^n \frac{1}{(2\pi)^6 V^2} \frac{1}{k_i^0 k_j^0} \Big(k^{iI} k_{jI}\Big).
 \end{split}}
\end{equation}

\section{Calculation of $\left\langle J^I_5 J_{I5}\right\rangle$}\label{appd}
It is quite evident that if we follow the same calculation for $J_{I5}$ as for the $J_I$, we have,
\begin{equation}
{\begin{split}
 \left\langle J^{I}_5 J_{I5}\right\rangle= -\sum_{j=1}^n \left(({S}^\mathrm{SA}_5)_{jj} + \sum_{\substack{l=1\\ l \neq j}}^n ({S}^\mathrm{SA}_5)_{jl}\right)
 \end{split}}
\end{equation}
where
\begin{equation}
  \begin{split}
 ({S}^\mathrm{SA}_5)_{jj} =& \frac{1}{2}\sum_{s,s_j} \int d^3k  \frac{1}{(2\pi)^6 V k^0 k_j^0} \big[ \Bar{u}(k_j, s_j) \gamma^I\gamma_5 u(k, s) \big] \big[ \Bar{u}(k, s) \gamma_I \gamma_5 u(k_j, s_j)\big]\\ &-\frac{1}{4}\sum_{s_i, s_j} \sum_{\substack{i=1 \\ i \neq j}}^n  \frac{1}{(2\pi)^6 V^2 k_i^0 k_j^0} \big[ \Bar{u}(k_j, s_j) \gamma^I \gamma_5 u(k_i, s_i) \big] \big[ \Bar{u}(k_i, s_i) \gamma_I \gamma_5 u(k_j, s_j)\big],
  \end{split}
\end{equation}
and
\begin{equation}
  ({S}^\mathrm{SA}_5)_{jl} (l\neq j)= \frac{1}{4}\sum_{s_i, s_j} \frac{1}{(2\pi)^6 V^2 k_l^0 k_j^0} \big[ \Bar{u}(k_l, s_l) \gamma^I \gamma_5 u(k_l, s_l) \big] \big[ \Bar{u}(k_j, s_j) \gamma_I \gamma_5 u(k_j, s_j)\big].
\end{equation}
In this case,
\begin{equation}
\begin{split}
 &\frac{1}{2} \sum_{s, s_j} \big[ \Bar{u}(k_j, s_j) \gamma^{I} \gamma_5 u(k, s) \big] \big[ \Bar{u}(k, s) \gamma_{I} \gamma_5 u(k_j, s_j)\big] = - \Big( 2 m^2 -  k^I k_{jI} \Big),\\
 &\frac{1}{4} \sum_{s_i,s_j} \big[ \Bar{u}(k_j, s_j) \gamma^{I} \gamma_5 u(k_i, s_i) \big] \big[ \Bar{u}(k_i, s_i) \gamma_{I} \gamma_5 u(k_j, s_j)\big] = - \frac{1}{2}\Big( 2m^2 -  k_i^{I} k_{jI} \Big),\\
 &\frac{1}{4} \sum_{s_i,s_j} \big[ \Bar{u}(k_i, s_i) \gamma^I \gamma_5 u(k_i, s_i) \big] \big[ \Bar{u}(k_j, s_j) \gamma_I \gamma_5 u(k_j, s_j)\big] = 0.
\end{split}
\end{equation}
Therefore,
\begin{equation}
{\begin{split}
 \left\langle J^{I}_5 J_{I5}\right\rangle=  \sum_{j=1}^n \int \frac{d^3k}{V} \frac{m^2}{(2\pi)^6} \frac{1}{k^0 k_j^0} \Big( 2 - \frac{1}{m^2} k^I k_{jI}\Big) - \frac{1}{2} \sum_{\substack{i, j=1 \\ i\neq j}}^n \frac{m^2}{(2\pi)^6 V^2} \frac{1}{k_i^0 k_j^0} \Big( 2 - \frac{1}{m^2} k_i^I k_{jI}\Big)\,.
 \end{split}}
\end{equation}.
\vspace{-0.9cm}
\section{Calculation of $\left\langle J^{I5}J_{I}\right\rangle$}\label{appcross}

Finally, for the cross current part we have,
\begin{equation}\label{axial1}
 \begin{split}
  \left\langle J^{I5}J_{I}\right\rangle &=- \frac{1}{2}\sum_{j=1}^n \sum_{s, s_j} \int d^3k  \frac{1}{(2\pi)^6 V k^0 k_j^0} \big[ \Bar{u}(k_j, s_j) \gamma^{I}\gamma_5 u(k, s) \big] \big[\Bar{u}(k, s) \gamma_{I} u(k_j, s_j)\big] \\
  &+\frac{1}{4} \sum_{\substack{i, j=1 \\ i \neq j}}^n \sum_{s_i, s_j}  \frac{1}{(2\pi)^6 V^2 k_i^0 k_j^0} \big[ \Bar{u}(k_j, s_j) \gamma^{I}\gamma_5 u(k_i, s_i) \big] \big[ \Bar{u}(k_i, s_i) \gamma_{I} u(k_j, s_j)\big] \\
  &- \frac{1}{4} \sum_{\substack{i, j=1 \\ i \neq j}}^n \sum_{s_i, s_j}\frac{1}{(2\pi)^6 V^2 k_i^0 k_j^0} \big[ \Bar{u}(k_i, s_i) \gamma^{I}\gamma_5 u(k_i, s_i) \big] \big[ \Bar{u}(k_j, s_j) \gamma_{I} u(k_j, s_j)\big].
 \end{split}
\end{equation}
If we carry out the involved spin sums the above expression \eqref{axial1} vanishes.

\section{Calculation of $\left\langle T_{\mu\nu}^\mathrm{free}\right\rangle$}
For a single species of fermions, using Eq. \eqref{solpsi} in Eq. \eqref{freeEMT} we get,
\begin{equation}
\begin{split}
 T_{\mu\nu}^\mathrm{free}\ket{n}=\frac{1}{4}\sum_{s,s^\prime} \int d^3k~  d^3k' \frac{i}{(2\pi)^3 V\sqrt{k^0 {k^\prime}^0}}e^{-i(k^\prime-k).x} \Big[\left(k_\mu+k^\prime_\mu\right) \Bar{u}(k', s')\gamma_\nu u(k,s)+(\mu\rightarrow\nu)\Big]\\ c^\dagger(\bm{k'}, s')c(\bm{k}, s)\prod_{i=1}^n c^\dagger(\bm{k_i}, s_i)\ket{0}.
 \end{split}
\end{equation}
Thus, using Eq. \eqref{1stmat}, we can write,
\begin{equation}
\begin{split}
 T_{\mu\nu}^\mathrm{free}\ket{n}=\frac{1}{4}\sum_{j=1}^n \sum_s\int d^3 k \frac{i}{(2\pi)^3 V\sqrt{k^0 {k_j}^0}}e^{-i(k-k_j).x}\Big[\left(k_\mu+k_{j\mu}\right) \Bar{u}(k, s)\gamma_\nu u(k_j, s_j)+(\mu\rightarrow\nu)\Big]\\{\prod_{l=1}^n}^\prime c^\dagger(\bm{k_l}, s_l)\ket{0},
 \end{split}
\end{equation}
which implies,
\begin{equation}
\begin{split}
\left\langle T_{\mu\nu}^\mathrm{free}\right\rangle=\bra{n} T_{\mu\nu}^\mathrm{free}\ket{n}=\frac{1}{4}\sum_{j=1}^n \frac{i}{(2\pi)^3 V {k_j^0}}\Big[2k_{j\mu} \Bar{u}(k_j, s_j)\gamma_\nu u(k_j, s_j)+(\mu\rightarrow\nu)\Big].
 \end{split}
\end{equation}
After taking the spin average we have,
\begin{equation}
\begin{split}
\left\langle T_{\mu\nu}^\mathrm{free}\right\rangle=\bra{n} T_{\mu\nu}^\mathrm{free}\ket{n}=\frac{2}{(2\pi)^3 V}\sum_{j=1}^n \frac{k_{j \mu} k_{j \nu}}{k_j^0},
 \end{split}
\end{equation}
from which Eq. \eqref{freeexp} follows as we consider multiple species of fermions and also take the large $n$-limit. 
\renewcommand{\theequation}{\thesection.\arabic{equation}}

\bibliography{ref}

\end{document}